\documentclass[12pt]{article}
\usepackage{amsfonts}
\normalbaselines
\catcode `\@=11
\@addtoreset{equation}{section}

\def\section{\@startsection {section}{1}{\z@}{-3.5ex plus -1ex minus
     -.2ex}{2.3ex plus .2ex}{\normalsize\bf}}

\def\subsection{\@startsection{subsection}{2}{\z@}{-3.25ex plus -1ex minus
 -.2ex}{1.5ex plus .2ex}{\normalsize\bf}}
\def\thebibliography#1{\section*{References\markboth
  {REFERENCES}{REFERENCES}}\list
  {[\arabic{enumi}]}{\settowidth\labelwidth{[#1]}\leftmargin\labelwidth
  \advance\leftmargin\labelsep
  \usecounter{enumi}}
  \def\newblock{\hskip .11em plus .33em minus -.07em}
  \sloppy
  \sfcode`\.=1000\relax}
 
\catcode `\@=12
\newcommand\ie{that is,\ }

\def\D{{D}}
\def\G{{\mathbb G}}
\def\L2{{\rm L^2}({\mathbb R}^2)}
\def\LT2{{\rm L^2}({\mathbb T}^{2})}
\def\LR2{{\rm L^2}({\mathbb R}^{2})}
\def\one{{\mathbf 1}}
\def\R2{{\mathbb R}^2} 
\def\T2{{\mathbb T}^2}
\def\Z2{{\mathbb{Z}}^2}%

\newcommand\delby[1]{\frac{\partial}{\partial #1}}
\newcommand\ep[1]{{\rm e}^{\textstyle #1}}
\newcommand\half{{\frac 1 2}}

\newcommand\bcdot{\bullet}
\newcommand\bull{\item}
\newcommand\eq[1]{(\ref{#1})}

\newcommand\ba{\begin{array}}
\newcommand\ea{\end{array}}
\newcommand\be{\begin{enumerate}}  
\newcommand\ee{\end{enumerate}}  
\newcommand\bi{\begin{itemize}}  
\newcommand\ei{\end{itemize}}  
\newcommand\bd{\begin{description}}  
\newcommand\ed{\end{description}}  
\newcommand\beq{\begin{equation}}  
\newcommand\eeq{\end{equation}}  
\newcommand\beqa{\begin{eqnarray}}  
\newcommand\eeqa{\end{eqnarray}}  

\begin{document}
\vspace*{2.5cm}
\noindent
{\bf PATH INTEGRAL QUANTIZATION FOR A TOROIDAL PHASE SPACE}\vspace{1.3cm}\\

\noindent

\hspace*{1in}

%%%%%%%%%%%%%%%%%%%%%%%%
%   Use this for a multi-author contribution
%%%%%%%%%%%%%%%%%%%%%%%%

\begin{minipage}{13cm}
Bernhard G.\ Bodmann and John R.\ Klauder\footnotemark \vspace{0.3cm}\\
Department of Mathematics, University of Florida,\\ 
Gainesville, FL 32611, USA \\
E-mails:  bgb@math.ufl.edu, klauder@phys.ufl.edu 
\end{minipage}\footnotetext{Also at the Department of Physics.}

\vspace*{0.5cm}
\begin{abstract}
\noindent
A Wiener-regularized path integral is presented as an alternative way
to formulate Berezin-Toeplitz quantization on a toroidal phase space.
Essential to the result
is that this quantization prescription for the torus 
can be constructed as an induced representation from anti-Wick quantization 
on its covering space, the plane. 
When this construction is expressed in the form of a Wiener-regularized 
path integral, symmetrization prescriptions for the propagator 
emerge similar to earlier path-integral formulas on multiply-connected 
configuration spaces.
\end{abstract}
%
% section 1

\section{\hspace{-4mm}.\hspace{2mm}INTRODUCTION}

With the notion of ``quantization'' we associate the construction of
quantum systems that are in correspondence with a given classical
system.  The various quantization prescriptions usually specify some
Hilbert space and a mapping of suitable classical observables to
operators on this Hilbert space. In Schr\"odinger's prescription for
canonical quantization, the vectors in the Hilbert space are
square-integrable functions on classical configuration space, and
their dynamics is derived from a partial differential equation
commonly known as Schr\"odinger's equation. There are other
quantization schemes in which the Hilbert space consists of functions
on the classical phase space, and among these, coherent states often
play an important role \cite{Kla63,Ber74,CGR90}.

An alternative way to express quantization is by path integration.
Thanks to the Feynman-Kac formula \cite{Sim79}, Schr\"odinger's 
approach can be associated with a Wiener integral over paths in
configuration-space.  Similarly, a formula of Daubechies and Klauder
\cite{DaKl85} relates anti-Wick quantization to a so-called 
Wiener-regularized
path integral, \ie a limit of certain Wiener-integrals over paths in
phase space.  In both approaches, only continuous paths contribute,
which raises the question how the quantization of classical systems
with topologically nontrivial phase spaces appears in these
formulations.  The torus as a simple example has been useful to
develop such concepts, and we will restrict ourselves to that case.

The main goal of this paper is to derive a Wiener-regularized path
integral formula associated to Berezin-Toeplitz quantization on a
toroidal phase space.  The result exhibits similarities to earlier
investigations of Feynman path integrals on multiply connected
configuration spaces \cite{Sch68,LaDW71,Dow72}.  The similar nature of
both results is traced back to the appearance of induced
representations which is a natural concept for the
quantization on quotient spaces \cite{Mac68,MMT95}.

\section{\hspace{-4mm}.\hspace{2mm}ANTI-WICK QUANTIZATION}
First we will quickly review how Berezin-Toeplitz quantization
applies to a system with a
canonical degree of freedom, \ie when the phase space can be
identified with the Euclidean plane $\R2$ equipped with the
standard symplectic structure.  In this special case Berezin-Toeplitz
quantization is also known as anti-Wick quantization.

\subsection{\hspace{-5mm}.\hspace{2mm}Operator Formulation}

The Hilbert space for the description of quantum mechanical states is 
chosen as a closed  subspace 
$\cal H$ of $\L2$ consisting of functions $\psi$ 
which are square integrable with respect to the two-dimensional
Lebesgue measure $dpdq$ on $\R2$ and in addition to that
satisfy the integral equation
\begin{equation}
   \label{eq:RepPty}
    \psi(p',q') = (K\psi)(p',q') 
        := 
      \int_{\R2} \frac{dp dq}{2 \pi \hbar} \, K(p',q';p,q) \, \psi(p,q)
\end{equation}
with Planck's constant $2 \pi \hbar$ and the so-called reproducing kernel
\begin{equation}
    \label{eq:RepKern}
    K(p',q';p,q) 
       := 
     \exp\Bigl\{-\frac{1}{4\hbar}\left[(p'-p)^2 + (q'-q)^2\right] +   
                  \frac i {2\hbar}   (pq'-p'q)\Bigr\}   \, .
\end{equation}

This specifies the kinematical background of the quantum theory.  The
transition from classical to quantum dynamics is achieved by a mapping
taking classical observables, represented by sufficiently nice
real-valued functions on the phase space, to self-adjoint operators on
$\cal H$. Given the classical Hamilton function 
$h: \R2 \rightarrow \mathbb R$, anti-Wick
quantization defines the quantum Hamiltonian by
\begin{equation}
  \label{eq:htoH}
     H\psi := K(h \psi)
\end{equation}
on functions $\psi$ from a domain that is a suitable extension of the 
domain of $h$ interpreted as multiplication operator on $\cal H$.

\pagebreak[3] Three remarks apply:
\bi
\bull
 It is well-known that this construction satisfies a correspondence 
 principle \cite{Ber74}.
\bull The above prescription can also be obtained from a group theoretical 
 approach. In fact,
 the reproducing kernel can be identified with the inner product of 
 so-called coherent vectors in $\cal H$, 
 \begin{equation}
  \label{eq:GNS}
   K(p',q';p,q) = (\eta_{p',q'}, \eta_{p,q}) ,
 \end{equation}
 which are given as the image $\eta_{p,q} := \D(p,q) \eta$
 of a reference vector
 $\eta(p,q) := K(p,q;0,0)$ under the unitary operators $\D(p,q)$ acting as
 a ray representation of the Heisenberg-Weyl group,
 \begin{equation}
  \label{eq:HWaction}
    \left(\D(p,q) \psi\right)(p',q') := \ep{i(pq'-p'q)/2\hbar}\psi(p+p',q+q')
 \end{equation}
 on vectors $\psi \in \cal H$. 
 This implies that the self-adjoint operators for position and momentum 
 defined as 
 the generators $Q := (- i \hbar \delby{p} \D)(0,0)$ 
 and $P := (i \hbar  \delby{q} \D)(0,0)$ 
 satisfy the canonical commutation relation $P Q - Q P = - i \hbar$.
\bull
 For the convenience of relating the reproducing kernel to the unitary 
 ray representation 
 $\D(p,q)$ we chose the space $\cal H$ and not the so-called Fock-Bargmann 
 space of holomorphic functions which is usually associated with 
 Berezin-Toeplitz quantization \cite{Ber74} for the planar phase space $\R2$. 
 The difference simply 
 amounts to redistributing a Gaussian weight from the functions to the inner 
 product \cite{Bar61}.
\ei

\subsection{\hspace{-5mm}.\hspace{2mm}Path Integral Formulation}

The stochastic process underlying the path
integrals considered hereafter is the Brownian bridge \cite{Sim79,Roe94}. 
Its realizations are continuous paths $b: [0,r] \mapsto \R2$
in the plane, starting at $b(0) = (p,q)$ and stopping at $b(r) = (p',q')$
after a time $r>0$.
As a Gaussian process it is fully characterized by its mean
and covariance,
\begin{eqnarray}
  \label{eq:mean}
       \langle b(s) \rangle &=& (p,q) + [(p',q') - (p,q)] s / r \\
  \label{eq:covariance}
       \langle b_j(s) b_k(u) \rangle - \langle b_j(s)  \rangle 
                                             \langle b_k(u) \rangle
            &=& \delta_{jk} \hbar^2 (\min\{s, u\} -  s u/ r ) \, ,
\end{eqnarray}
respectively, with  $s,u \in [0,r]$.
Kronecker's delta $\delta_{jk}$ shows that the path components denumerated 
with $j,k \in \{1, 2\}$ are independent of each other.
For notational convenience we also introduce the conditional Wiener 
measure $\mu_{p',q',r;p,q}$, which relates to the expectation with respect 
to the Brownian bridge according to
\begin{equation}  \label{eq:CondWienerExp}
   \int \! d \mu_{p',q',r;p,q} \, (\bcdot) 
              := 
    \frac{1}{2\pi \hbar^2 r} \ep{-((p'-p)^2 + (q' - q)^2) / 2 \hbar^2 r} 
                                        \langle (\bcdot) \rangle \, .
\end{equation}

It was demonstrated by Daubechies and Klauder \cite{DaKl85} that 
anti-Wick quantization can be expressed in a 
so-called Wiener-regularized path integral. 
More precisely, the continuous integral kernel 
of the time evolution operator $\ep{-it H/\hbar}$ 
constructed from the Hamiltonian \eq{eq:htoH} 
is expressed as a limit of conditional Wiener-integrals  
\beq  \label{eq:DK} 
 \left(\ep{-it H/\hbar}\right)(p',q';p,q) = 2 \pi \hbar \lim_{r \to \infty} 
    \ep{ r\hbar / 2} 
     \int d\mu_{p',q',r;p,q} \, \ep{i {\cal S}_r/\hbar}                  
\eeq       
with the action functional  
\begin{equation}  
   {\cal S}_r(b) 
      := 
     \half \int_0^r \left[db_2(s) b_1(s) - db_1(s) b_2(s)\right]  
           - \frac{t}{r}\int_0^r \! ds\, h(b(s))   \, .
\end{equation}
The stochastic integral appearing in ${\cal S}_r$ is 
understood in the sense of Stratonovich \cite{ReYo94}.      

\pagebreak[3] Two remarks apply:
\bi
\bull 
The above path integral expression is valid for all Hamilton functions $h$ 
that are
bounded from below and polynomially bounded from above \cite{DaKl85}.
\bull Setting $t = 0$ we obtain the reproducing kernel and thus all   
the information about the Hilbert space $\cal H$. Clearly, the path integral 
comprises information about quantization for both kinematics and 
dynamics in a nutshell!   
%In particular, the Markov property in the Wiener integral representation 
%can be used to deduce the reproducing kernel property. 
\ei

\section{\hspace{-4mm}.\hspace{2mm}INDUCED QUANTIZATION FOR 
         A TOROIDAL PHASE SPACE}   

Let us regard the torus as the quotient space of the Euclidean plane $\R2$ 
where points are identified which can be mapped into each other by adding 
vectors $g$ from a grid
${\mathbb G}:=(a{\mathbb Z}, b{\mathbb Z})$  
of spacings $a, b > 0$. For convenience we will denote the points
on the torus with their representatives from the half-open rectangle 
$\T2 := [0,a)\times[0,b)$.  

Classically, Hamiltonian dynamics on the torus ${\mathbb T}^2$ can be
lifted to the covering space $\R2$ by 
periodically extending $h$ on $\R2$ to make it compatible with the
quotient construction.  In the following section we will implement
this concept within the framework of quantum mechanics.

\subsection{\hspace{-5mm}.\hspace{2mm}Operator Formulation}
It is well-known that Berezin-Toeplitz quantization only gives rise to
Hilbert spaces of nonzero dimension if the volume of the torus is
integral in units of Planck's constant \cite{CGR90,BHSS91},
\ie $ab=2 \pi \hbar N$ with a positive integer $N$.  

These Hilbert
spaces, denoted as ${\cal H}_k$, are realized here (with
an analogous identification as for $\cal H$ and Fock-Bargmann space)
as closed subspaces
of the space $\LT2$ of square-integrable functions on the torus.
Each function $\psi$ in ${\cal H}_k$ satisfies
\beq
    \psi(p',q') = \int_{{\mathbb T}^2} \frac{dp dq}{2 \pi \hbar} \,
                         K_k (p',q';p,q) \, \psi(p,q)
\eeq
analogous to \eq{eq:RepPty}, where
the reproducing kernel   
\begin{equation}
  \label{eq:RepkKern}  
    K_k(p',q';p,q)  := \sum_{g \in \G}  (\eta_{p',q'}, \D_{g,k} \eta_{p,q})   
\end{equation}   
is defined in terms of the inner product and the coherent vectors
of the preceding section and the unitary operators of the form
\begin{equation}
  \label{eq:Dgk}
      \D_{g,k} := \D(g_1,g_2) \, 
       \ep{i(g_1 k_2 -  g_2 k_1 + \half g_1 g_2)/\hbar} 
\end{equation}  
with a parameter $k$ chosen from the ``inverse'' 
torus $[0, \frac{2\pi\hbar}{a})\times [0, \frac{2\pi\hbar}{b})$.

\pagebreak[3] A few remarks apply: 
\bi
\bull
With the integrality of $ab/2\pi\hbar$ and the definition \eq{eq:HWaction}
it can be confirmed that the set of
operators $\{\D_{g,k}\}_{g \in \G}$ is commutative and forms a genuine
unitary group representation of the discrete phase-space translations
by grid vectors $g \in \G$. 

Thanks to the abelian group composition law for this set
and the reproducing kernel property of $K$ on $\cal H$
it follows that $K_k$ is also a reproducing kernel.

The additional freedom in the parameter $k$ 
amounts to choosing a character of $\G$, which 
reflects in the kernel and thus in the boundary conditions
that the functions in ${\cal H}_k$ obey.
\bull 
The Hilbert space ${\cal H}_k$ is $N$-dimensional. This can be 
read off from an orthogonal decomposition of the kernel
\begin{eqnarray}
  \label{eq:orthdecomp}
  K_k(p',q';p,q) &=&\sum_{j=0}^{N-1} \phi_j^*(p',q') \phi_j(p,q) \, ,\\
\phi_j(p,q) &:=& \Bigl(\frac{4\pi \hbar}{a^2}\Bigr)^{1/4} \, 
       \ep{-ipq/2\hbar} \, \ep{-i p(k_2 + b j / N)/\hbar}      \\
        & & \times \sum_{n \in \mathbb Z} \ep{ibn (p + k_1) / \hbar} \,
                    \ep{-  (q + k_2 + bj / N - bn)^2 / 2 \hbar} \nonumber
\end{eqnarray}
which also shows the connection to the (holomorphic) 
theta functions spanning the usual Hilbert space 
\cite{CGR90,LV90,BHSS91} obtained
from Berezin-Toeplitz quantization for the torus.
The decomposition \eq{eq:orthdecomp} 
is derived with the help of Poisson's sum formula, the   
use of the identity $\D_{-g,k} Q \D_{g,k} = Q + g_2\one$ and the   
fact $ab=2\pi \hbar N$ by rewriting the summation 
in \eq{eq:RepkKern} as
\begin{equation}   
   \sum_{g \in \G} \D_{g,k} = \frac{b}{N} \sum_{m,n\in \mathbb Z} 
     \sum_{j=0}^{N-1} \ep{-ib m (P+k_1)/\hbar} 
                         \, \delta(Q + k_2 + \frac{b}{N} j) \,    
                           \ep{i b n (P+k_1)/\hbar} \, . 
\end{equation} 
%The orthogonal decomposition also shows that
%the diagonal of the kernel $K_k(p,q;p,q)$ is manifestly positive as expected.
\bull
The definition of the reproducing kernel \eq{eq:RepkKern}
can be interpreted as
an induced representation. We briefly sketch this idea following the
presentation in \cite{Lan96}.  
By the preceding remark, 
$p_k: \psi \mapsto (\mkern2mu\sum_g (\psi, \D_{g,k} \psi)\mkern1mu )^{1/2}$
defines a semi-norm on $\cal H$. Passing to the quotient space of $\cal H$ 
with respect to the set ${\cal M}_k:= \{\psi: p_k(\psi) = 0\}$ of vectors 
having zero length and discarding vectors of infinite length yields 
a pre-Hilbert space equipped with the inner product derived from $p_k$. 
The completion of this space is isomorphic to ${\cal H}_k$. 
In the above setting, the symmetrization
$(\psi+{\cal M}_k)(p,q) := \sum_g (\D_{g,k} \psi)(p,q)$ 
realizes the quotient mapping, and the inner product 
derived from $p_k$ coincides with that of $\LT2$.   
\ei

%\subsection{\hspace{-5mm}.\hspace{2mm}Quantization of Dynamics}
   
Once the reproducing kernel is given, it is straightforward to mimick the 
same procedure as for anti-Wick quantization and define the Hamiltonian as
\begin{equation}
     \label{eq:Hk}
      H_k \psi := K_k (h \psi) 
\end{equation}
for suitable $\psi \in {\cal H}_k$.  
Again it is known that a correspondence principle holds \cite{BHSS91,BMS94}
for this quantization prescription. 

Alternatively, the continuous integral kernel for the time evolution 
operator constructed from $H_k$ can
be directly expressed by a symmetrization analogous to \eq{eq:RepkKern},
\begin{equation}    \label{eq:SymKern}
     \left(\ep{-it H_k/\hbar}\right) (p',q';p,q) 
         = 
      \sum_{g \in \G} (\eta_{p'q'}  , 
                        \D_{g,k} \, \ep{-it H/\hbar}  \eta_{p,q})   \, .
\end{equation}   
Hereby $H$ is the operator obtained from anti-Wick quantization of the 
periodic extension of $h$ on $\R2$. 

We will establish \eq{eq:SymKern} under the assumption that $h$ is bounded.
It is sufficient to show that the right-hand side
defines an integral kernel of the one-parameter group of unitary operators
$\ep{-it H_k/\hbar}$ and that it is continuous in the parameters $p',q'$
and $p,q$. 
Unitarity and the group composition law are straightforward to verify.
By the definition \eq{eq:Hk}, 
the generator of this group can then be identified as $H_k$.
To see the claimed continuity, one 
shows that the expansion 
\begin{equation}
  \label{eq:uniform}
    \sum_{g \in \G} (\eta_{p'q'}  , 
                        \D_{g,k} \, \ep{-it H/\hbar}  \eta_{p,q}) 
  =
    \sum_{n=0}^\infty \frac{(-it)^n}{\hbar^n n!} 
     \sum_{g \in \G} (\eta_{p',q'}, \D_{g,k} H^n \eta_{p,q}) 
\end{equation}
provides a uniformly convergent series. 
This follows from H\"older's inequality, a bound on the functions
$\eta_{p',q'} + {\cal M}_k$, and an estimate on the norms of 
$ H^n \eta_{p,q}$ in the Banach space ${\rm L}^1(\R2)$ of 
Lebesgue-integrable 
functions on the plane. The last estimate is obtained from
the boundedness of $h$ interpreted as a multiplication operator
on ${\rm L}^1(\R2)$ and the boundedness of the operator on 
${\rm L}^1(\R2)$ associated with the kernel $K(p',q';p,q)$. 
 
\subsection{\hspace{-5mm}.\hspace{2mm}Path Integral Formulation} 
 
The quantization prescription 
in the preceding section
can be expressed in the path integral 
formula   
\beq   \label{eq:DKT}
 \left(\ep{-it H_k/\hbar}\right)(p',q';p,q) = 2 \pi \hbar \lim_{r \to \infty} 
             \ep{r\hbar / 2} 
              \sum_{g \in \mathbb G} 
               \ep{i f_{g,k}(p',q')/\hbar}        
  \int \! d\mu_{p'+g_1, q'+g_2,r;p,q} \, \ep{i{\cal S}_r/\hbar}    
\eeq      
with the function 
$ f_{g,k}(p',q'):= g_1 k_2 - g_2 k_1 + \half (g_1 q' - g_2 p' + g_1 g_2) $
and the previous action functional ${\cal S}_r(b)$ containing the periodic 
extension of $h$. 
This result is derived from \eq{eq:SymKern} in combination 
with the Daubechies-Klauder formula \eq{eq:DK} under the
assumption that $h$ is bounded. 

\pagebreak[3] We conclude with two remarks:
\bi
\bull 
 The general structure of the path integral formula \eq{eq:DKT}
 is analogous to the construction in \cite{Sch68}.
 However, the extra phase factors independent of $k$ cannot be anticipated by
 the argument given there, which is 
 only concerned with multiply-connected configuration spaces.  
\bull 
 With hindsight one can interpret the result as an integral over
 the actual paths on the torus.  This shows that each path receives a
 phase factor depending on its homotopy class and the transition
 functions of the complex line bundle underlying Berezin-Toeplitz
 quantization.  From this viewpoint one may hope to deduce
 generalizations for more general phase-space manifolds.  
\ei
  
{\noindent {\bf Acknowledgement\ }
We would like to thank the organizers 
for creating such a lively and stimulating workshop.}

%%%%%%%%%%% Insert your bibliography below %%%%%%%%%%

\end{document}